\documentclass[twocolumn,showpacs,preprintnumbers,amsmath,amssymb]{revtex4}
\usepackage{tabularx,graphicx}

\usepackage{color}
\usepackage{hyperref}
\hypersetup{
    colorlinks=true,
    linkcolor=blue,
    filecolor=blue,      
    urlcolor=blue,
}

\usepackage{color}

\usepackage{ulem}   

\begin{document}

\newcommand{\beq}{\begin{equation}}
\newcommand{\eeq}{\end{equation}}
\newcommand{\beqn}{\begin{eqnarray}}
\newcommand{\eeqn}{\end{eqnarray}}
\newcommand{\bmath}{\begin{subequations}}
\newcommand{\emath}{\end{subequations}}
\newcommand{\bra}[1]{\langle #1|}
\newcommand{\ket}[1]{|#1\rangle}

\title{Understanding electron-doped cuprate superconductors as hole superconductors}

\author{J. E. Hirsch$^{a}$  and F. Marsiglio$^{b}$ }
\address{$^{a}$Department of Physics, University of California, San Diego,
La Jolla, CA 92093-0319\\
$^{b}$Department of Physics, University of Alberta, Edmonton,
Alberta, Canada T6G 2E1}

\begin{abstract} 
Since their experimental discovery in 1989, the electron-doped cuprate superconductors have presented both a major challenge and a major opportunity.
The major challenge has been to determine whether these materials are fundamentally different from or essentially similar to their hole-doped counterparts;
a major opportunity  because answering this question would strongly constrain the possible explanations for what is the essential physics that leads to  high temperature 
superconductivity in the cuprates, which is still not agreed upon. Here we argue that   experimental results  over the past 30 years
on electron-doped cuprate materials   have    provided conclusive answers to  these fundamental questions, by establishing that both in hole- and electron-doped cuprates, superconductivity originates in
pairing of  hole carriers in the same band. We discuss a model to describe this physics that is different from the generally accepted ones, and calculate  physical observables that agree with experiment, in particular tunneling characteristics. We
 argue that our model is simpler, more natural and more compelling than   other models. Unlike other models, ours was originally proposed before rather than after many key experiments were performed.
\end{abstract}
\maketitle 

\section{introduction}

  Shortly after the discovery of high temperature superconductivity in cuprate superconductors in 1986 \cite{alex}, it became clear 
  that the carriers responsible for superconductivity in these materials were $holes$ \cite{h1,h2,h3,h4,h5,h6}. Upon changing the chemical composition of the parent insulating compound
  so that hole carriers were added to the copper-oxygen planes, the superconducting $T_c$  was found  to increase, go through a maximum and then decrease to zero in the
  `overdoped' regime \cite{h3,torrance}. 
  
Then, on  January  26, 1989 it was  reported   by Takagi, Tokura and Uchida \cite{edoped} that by doping a parent insulating material  with {\it   electrons} instead of holes 
superconductivity also occurred, albeit with a 
smaller maximum critical temperature. 
Initially, experiments appeared to show very definitely  that
indeed the  charge carriers in these electron-doped materials were electrons \cite{edoped,tranquada,tokura2}, as the title
of Ref. \cite{edoped} claimed: {\it ``A superconducting
copper oxide compound with electrons as the charge
carriers''}.
The general reaction to this discovery was that it provided evidence for an approximate electron-hole symmetry \cite{emery,rice,pool,maple}.
For example, Art Sleight stated in March 1989 \cite{sleight} ``{\it This symmetry between adding and subtracting electrons will have to be reflected in any theory that explains high-temperature superconductivity, and existing theories based on the supposition that there is something unique about hole carriers are `out the window'}.''

Electron-hole symmetry is to be expected if the undoped parent insulating compound is assumed to be  described by a half-filled band governed by the Hubbard Hamiltonian \cite{anderson}, which is particle-hole symmetric, and doping of holes or electrons results in changing the carrier occupation 
{\it in this band}. This
 was widely assumed to be the case at that time   and continues to be widely assumed to be the case today.
 
            \begin{figure}[]
 \resizebox{4.5cm}{!}{\includegraphics[width=6cm]{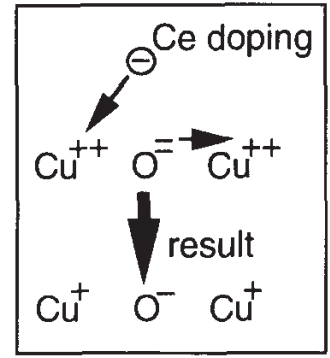}}
 \caption {  Schematic depiction of how holes are created by electron
doping. The electron added to $Cu^{2+}$ repels an electron from $O^{2-}$ to
the neighboring $Cu^{2+}$, leaving behind a hole in oxygen ($O^-$).
 }
 \label{figure1}
 \end{figure}

Instead,   immediately  after the discovery of the electron-doped  materials we pointed out  \cite{natureletter,tunn,mrs,carrier,m2s89} that a natural explanation existed for why hole carriers of the same nature as the hole carriers in the hole-doped materials \cite{tang} would be induced in the electron-doped materials, 
and   we predicted that 
subsequent experiments would show that  hole carriers exist and are  responsible for superconductivity also in the electron-doped materials \cite{natureletter,tunn,mrs,carrier,m2s89}. 
Figure 1 shows schematically how holes on $O^=$ can result from electron-doping of $Cu^{++}$, we will delve into the details later. No experimental evidence suggesting neither that hole carriers existed nor that they were responsible for superconductivity 
in these materials existed at that time.

Already very soon thereafter, EELS experiments suggested the presence of holes at the oxygen sites in electron-doped
cuprates \cite{nucker}.   That  holes participate in the transport  was shown by detailed and extensive magnetotransport measurements
by several different experimental groups  extending over many  years \cite{wang,fortune,crusellas,suzuki,greene1,greene2,gollnik,greene3,greene4,greene5,greven}.
These experimental results and their analysis   showed that there are both electron and hole charge carriers in the electron-doped cuprates,  that hole carriers dominate the transport in the
regime where electron-doped cuprates become superconducting, and that it is the hole carriers that  likely drive superconductivity in these materials \cite{wang,greene5,greven}.  

However  several key questions remain unsettled. What is the nature of the electron and hole carriers in the electron-doped materials? Why, if there are hole carriers in electron-doped cuprates,
aren't there electron carriers in hole-doped cuprates?
Are the hole carriers in the electron-doped materials  of the same nature as those  in the hole-doped materials?  
Even if they are, is the pairing mechanism the same? We argue that definite answers to these questions
whould go a long way towards elucidating the origin of superconductivity in both hole- and electron-doped cuprates.

In this paper we  elaborate on the simple answers to these questions that we proposed 30 years ago 
\cite{natureletter,tunn,mrs,carrier,m2s89}, 
and argue that various experimental results obtained during these 30 years support our original proposal. We will also argue that 
other proposals to explain these questions are complicated, unnatural and implausible. 

In a nutshell, our proposal was and is: hole carriers responsible for superconductivity in $both$ hole-doped and electron-doped
materials reside in a band resulting from overlapping oxygen $p\pi$ orbitals in the Cu-O$_2$ plane that point perpendicular to the $Cu-O$ bonds,
as shown in Fig. 2. This band is
full in the undoped case and becomes slightly less than full both on the hole-doped and the electron-doped side, for
reasons we will explain.
The electron carriers in the electron-doped cuprates reside in the $Cu-O$ band formed by the overlapping
$Cu$  $d_{x^2-y^2}$ and $O$ $p\sigma$ orbitals pointing along the $Cu-O$ bond. Proposals that hole carriers in the 
hole-doped cuprates reside in the O $p\pi$ orbitals were also made early on by Goddard et al \cite{goddard}, 
Stechel and Jennison \cite{sj}, 
Birgeneau et al \cite{birg} and Ikeda \cite{ikeda}.

          \begin{figure}[]
 \resizebox{6.5cm}{!}{\includegraphics[width=6cm]{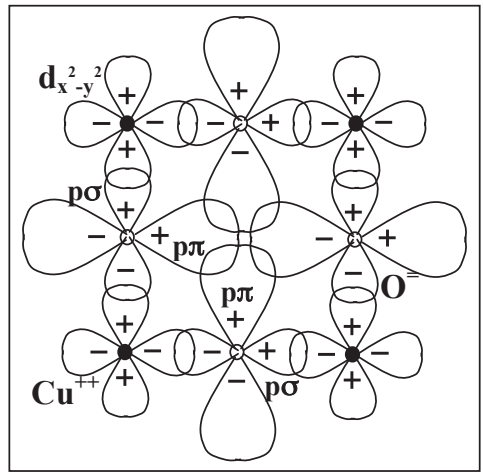}}
 \caption {  $Cu$  $d_{x^2-y^2}$ and oxygen orbitals in the Cu-O planes. In the undoped parent compound the nominal valence is $Cu^{++}$ and
 $O^=$ and there is one hole in the filled Cu $d^{10}$  orbital. The O $p\pi$ orbitals point perpendicular to the
 Cu-O bonds, the $p\sigma$ orbitals parallel. We propose that doped holes reside in a band resulting principally from overlapping 
 O $p\pi$ orbitals for both hole- and electron-doped cuprates.
 }
 \label{figure1}
 \end{figure} 

\section{ two-band models}
As discussed above, experiments indicate    \cite{wang,fortune,crusellas,suzuki,greene1,greene2,gollnik,greene3,greene4,greene5,greven}
that hole carriers exist and are responsible for superconductivity
in both electron-doped and hole-doped cuprate superconductors. Furthermore, these experiments show
that in electron-doped cuprates there is two-band conduction
in the normal state, with the other band being electron-like. The question is, where are these carriers? We start by giving a brief overview of three different possibilities that have been proposed: (i) Our two-band model, (ii) two-band $t-J$ model, and
(iii) reconstructed Fermi surface models.

\subsection*{(i) Our two-band model}
The simplest way to have two-band conduction for the system shown in Fig. 2 is if one band
involves principally the O $p\sigma$ orbitals
hybridized with the $Cu$  $d_{x^2-y^2}$, which we will call the Cu-O band,  and the other band 
involves principally
the O $p\pi$ orbitals orbitals pointing perpendicular to the  O $p\sigma$ orbitals in the plane, which we will call the O   band. For the hole-doped materials we had proposed \cite{tang,marsiglio1}, 
before the electron-doped materials were discovered, that their
high $T_c$ could be understood as arising from hole carriers in the O band.
Furthermore within this theoretical framework it is predicted that superconductivity can $only$ arise from hole carriers in a 
nearly full band \cite{hole1989,hole1990,holesc}.  In other words,  within this theoretical framework {\it `A superconducting copper oxide compound with electrons as the charge
carriers}', as announced \cite{edoped} by Tokura, Takagi and Uchida on January 26, 1989, 
  cannot exist. As we now know from experiments \cite{ wang,fortune,crusellas,suzuki,greene1,greene2,gollnik,greene3,greene4,greene5,greven},
 it does not exist, at least to date.

There is a simple way to understand why doping with electrons can create holes in the O band, illustrated in Fig. 3, in the hole representation. 
First, we assume the O $p\pi$ energy level for a hole is lower than the O $p\sigma$ level, in other words it costs less energy to remove an
electron from the O $p\pi$ orbital than from the O $p\sigma$ orbital. This is plausible for two reasons: first, as pointed out by
Birgeneau et al \cite{birg} and Goddard and coworkers \cite{goddard}, because there is more negative charge near the
center of a plaquette than along the Cu-O-Cu line, it costs less Coulomb energy to remove an electron (create a hole) from the
$p\pi$ orbitals that point towards the center of the plaquette than from the $p\sigma$ orbitals directed along the Cu-O bond.
Second, as discussed in ref. \cite{cuo}, the orbital relaxation effect that occurs when an electron is removed from the
$O^=$ ion is stronger if the electron is in the $p\pi$ orbital that is doubly occupied (by electrons) in the undoped case  than if it is in the $p\sigma$ orbital that is
only about $1.5$ occupied because of its hybridization with the neighboring Cu atoms. That lowers the energy for creating a
hole in the $p\pi$ orbital relative to creating it in the $p\sigma$ orbital.

           \begin{figure}[h]
 \resizebox{6.5cm}{!}{\includegraphics[width=6cm]{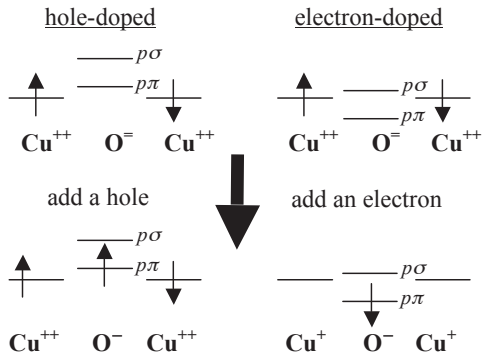}}
 \caption { Illustration of how oxygen hole carriers are created by both hole and electron doping in 
 high $T_c$ cuprates  in the hole representation (see text). Arrows denote holes. The difference in the relative locations of the 
 $O^=$ and $Cu^{++}$ orbitals in the hole-doped and electron-doped cases arises due to their different
 crystal structures, T and T$^\prime$.
 }
 \label{figure1}
 \end{figure}
 
In the undoped case, there is one hole at each $Cu^{++}$ site. For the electron-doped material,
we assume the single-hole O $p\pi$ energy level is lower than the $Cu$ energy level. Nevertheless in the undoped compound 
 the Cu hole doesn't `fall' onto the neighboring $O^=$ ion because of the cost in  Coulomb
repulsion between neighboring holes at Cu and O sites. Upon electron doping  the hole is removed from (electron is added to) a $Cu^{++}$ ion, now the hole from a neighboring $Cu^{++}$  can  fall into the $O^=$ ion without paying nearest neighbor Coulomb repulsion
energy, as illustrated in the lower right panel of Fig. 3. The net result from adding an electron to a $Cu^{++}$
is two $Cu^{+}$ ions and one $O^-$ ion, with the  hole in the $p\pi$ orbital. In other words, two extra electrons  {reside} in the Cu-O band and
one hole   in the O band.

This process can (and will) happen in the electron-doped cuprate materials because of the absence of
apical oxygens in their structure (T$^\prime$ structure), that increases the electrostatic potential at the 
$Cu^{++}$ site relative to the case of the hole-doped materials (T structure). Because the apical oxygen is relatively closer to the
$Cu$ atoms than to the  the $O$ atoms in the plane, the $Cu$ hole energy level is
relatively higher with respect to the $O$ levels in the T$^\prime$  structure (right side in Fig. 3). This facilitates $both$
the electron-doping of the material (it is not possible to dope electrons in the T structure) $and$
the transfer of electrons from $O^=$ sites to neighboring $Cu^{++}$ sites, thus creating O holes.
A detailed analysis of the energetics of these processes and the important role of reduction in 
getting the carriers to delocalize is given in ref.  \cite{reduction}.

In all the other models that have been proposed to describe the electron-doped and hole-doped cuprate superconductors
 in recent years, the O $p\pi$ orbitals are not included. It is assumed that the O  band is far below the
 Fermi energy and can be ignored. We discussed in Ref. \cite{cuo} why this may not be so.

\subsection*{(ii) Two-band $t-J$ model}

Another possible model to give rise to two-band superconductivity with electrons and holes was suggested by T. Xiang and coworkers \cite{2bandshen}.
They proposed that the two bands in question are a Zhang-Rice singlet band and the upper Hubbard band,
both originating in the overlap of orbitals $Cu d_{x^2-y^2}$ and $O p\sigma$ shown in Fig. 2 (what we called the Cu-O band).
The authors argue that both the Zhang-Rice singlet band and the upper Hubbard band should be described by
effective one-band $t-J$ models, or, more accurately, a  `hybridized two-band t-J model'. Then, this model maps onto
a one-band $t-U-J$ model, using approximations that according to the authors `may not
be fully satisfied in real materials'. The authors argue that this model gives results for a Fermi surface density map
consistent with ARPES observations. However, they don't explain how this model explains 
the transport experiments \cite{wang,fortune,crusellas,suzuki,greene1,greene2,gollnik,greene3,greene4,greene5,greven}
that
clearly show electrons and hole carriers in two different bands.

\subsection*{(iii) Reconstructed Fermi surface models}
In these models it is suggested that some kind of translational symmetry breaking with wavevector $(\pi,\pi)$ doubles the unit cell and this gives rise to
both electron and hole carriers. Lin and Millis \cite{millis} argue that this reconstruction of the Fermi surface occurs below Ce concentration 
$x=x_c=0.16$  due to antiferromagnetic long-range order, giving rise to electron pockets at $(0,\pi)$ and small hole pockets at 
$(\pi/2,\pi/2)$, while for $x >  x_c$ only a large hole-like Fermi surface exists occupying about half the Brillouin zone.
However, Motoyama and coworkers report \cite{moto} that the antiferromagnetic order disappears already at $x=0.134$, and that in the doping regime where electron-doped cuprates superconduct, only
short range antiferromagnetic correlation exists. This is incompatible with the hypothesis that the hole carriers responsible
for superconductivity arise from reconstruction of the Fermi surface due to antiferromagnetic order.
While it has been speculated that Fermi surface reconstruction may still occur in the electron-doped materials at high doping \cite{helm2011,helm2015,higgins2018} due to some other mechanism, e.g. 
a hidden d-density wave order \cite{chak}, no experimental evidence for this has been found so far.

We argue that our proposed model (i) is simpler and more natural than models (ii) and (iii). In the following sections we discuss in more detail experimental and
theoretical reasons in favor of our model versus the other models.
 \newline

 \section{magnetotransport}
 
  Initially, Hall coefficient measurements on electron-doped cuprates yielded a negative Hall 
 coeffcient \cite{tokura2}, consistent with the expectation that these were superconductors
 `with electrons as the charge carriers' \cite{edoped}. Later this changed, when experiments were performed
 on single crystals and thin films.
 
 Figure 4 shows measurements of the Hall coefficient versus temperature for a range of doping levels for a typical 
 electron-doped material, $Pr_{2-x}Ce_xCuO_4$ \cite{greene3}, and a hole-doped material, $(La_{2-x}Sr_x)_2CuO_4$ \cite{hallhole}.
 Results for $T_c$ versus doping are also shown \cite{krock,hallhole}.  There is a clear difference in the behavior of the Hall coefficient $R_H$ in both cases.

            \begin{figure}[h]
 \resizebox{8.5cm}{!}{\includegraphics[width=6cm]{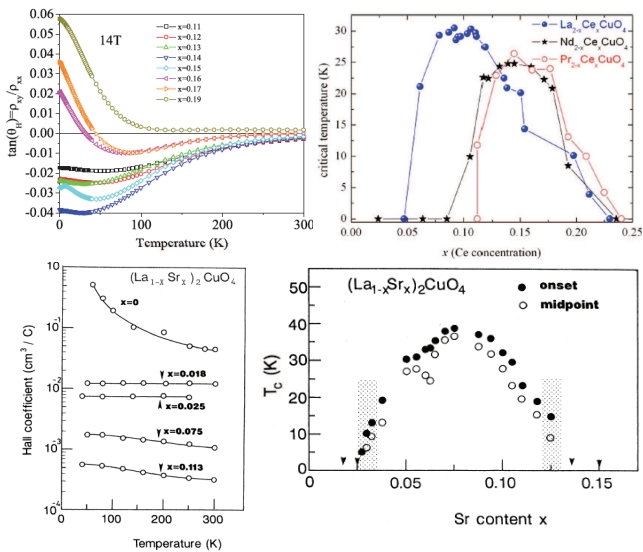}}
 \caption {Temperature dependence of Hall coefficient and $T_c$ versus doping for
 electron-doped (upper panels) \cite{greene3,krock} and hole-doped (lower panels) \cite{hallhole} cuprates.}.
 \label{figure1}
 \end{figure}
 
 In the hole-doped material, 
 $R_H$ is almost temperature independent and positive over the entire doping range where superconductivity
 occurs (up to $x=0.125$). For doping approaching $x=0.125$ the positive Hall coefficient becomes very small. 
 For doping $x=0.15$ and above (not shown), the Hall coefficient is negative and no superconductivity exists \cite{hallhole}. 
 
 These results for the Hall coefficient of hole-doped cuprates  were initially
  qualitatively  interpreted \cite{hallhole} as a crossover from a Mott-Hubbard regime ($R_H>0$) at low doping to a Fermi liquid regime ($R_H<0$) at high doping \cite{hallfl}  in a single
 band model.  However contrary to initial expectations it was found  within dynamical mean field theory that  the Hall coefficient of a 
 half-filled Hubbard model doped with holes is negative for all dopings 
 and given by the bare band structure results \cite{dmfthall,dmfthall2}. 
 At finite temperatures it was reportedly  found   that the Hall coefficient can   turn positive \cite{dmfthall2,dmfthall3,dmfthall4}; however the temperature and doping dependence does not
 resemble the experimental results shown in Fig. 4. In addition, these treatments predict that
 $R_H(\delta)=-R_H(-\delta)$, where $\delta$ is the doping away from half-filling, which is very different from the behavior
 shown in Fig. 4.
 
 Instead, we argue that the results for $R_H$ for hole-doped cuprates shown in Fig. 4, namely the near temperature
 independence and the positive value decreasing with hole doping, are most simply interpreted
 as resulting from doping of a single  band from an initial state where the band is completely full, as predicted by our model,
 where the band is the $O p\pi$ band.  As hole carriers are added the magnitude of the Hall coefficient decreases as expected
 from the simple formula $R_H\sim 1/(n_hec)$. As the band becomes half full   the Hall coefficient
 would change sign from positive to negative in a simple picture. The change in sign at high hole doping from positive to negative  before the band
 is half-full can be
 simply explained as due to a high scattering rate that would prevent carriers from
 completing closed hole orbits without scattering, which is necessary for the Hall coefficient to be positive.
 
  The behavior of the Hall coefficient for electron-doped cuprates as a function of temperature and doping shown in 
 Fig. 4 is drastically different from a `mirror image' of the hole-doped cuprates with a sign change, as would be predicted
 by  a  single band model. It shows unmistakable evidence for two-band conduction, one with electron carriers
 and one with hole carriers, with electrons dominating at low electron doping and holes dominating at high 
 electron doping. 
 For an isotropic two band model with electron and hole carriers of densities $n_e$ and $n_h$  the Hall coefficient is
\beq
R_H=-\frac{1}{n_e ec}\frac{1-(n_h/n_e)(\mu_h/\mu_e)^2}{1+(n_h/n_e)(\mu_h/\mu_e)^2}
\eeq
and will be negative if the mobility of the  hole carriers ($\mu_h$)  is much smaller than that of the electron carriers ($\mu_e$).  Wang et al  \cite{wang} and Crusellas et al  \cite{crusellas} have argued that the hole mobility increases rapidly as the temperature is lowered
due to a decrease in the hole scattering rate, and have
fitted the temperature and doping dependence of resistivity and Hall coefficient measured in experiments
 using reasonable assumptions for temperature-dependent electron and hole mobilities.  Wang et al pointed out \cite{wang} the similarity in the temperature dependence
 of the scattering rate of the hole carriers in hole-doped cuprates inferred from the resistivity with that of the hole carriers in electron-doped cuprates inferred from $R_H$, suggesting it is the same carriers and the same scattering processes. Crusellas et al pointed out 
  \cite{crusellas}
  that the measured hole mobility in electron-doped
 cuprates  is very similar to
that observed in hole-doped cuprates.
Note also  that in a single-band situation a temperature-dependent scattering rate does not
 result in temperature dependence of the Hall coefficient.

 Further evidence for the existence of two-band conduction in the regime where electron-doped cuprates superconduct comes from the sign and magnitude of the
 magnetoresistance. Already in 1994 Jiang et al
  pointed out \cite{greene1} that ``we see a remarkable correlation between
the occurrence of superconductivity and the appearance
of a positive MR'' and that a ``signature of two-band conduction is a positive
magnetoresistance''. Recently   Li et al \cite{greven}   pointed out that there is a considerable
 increase in the magnitude of the magnetoresistance in the region where bulk superconductivity
 is first seen, which reveals the underlying two bands. 
 Jiang et al \cite{greene1}  as well as Fournier et al \cite{greene2} also pointed out that the anomalously large Nerst coefficient they observed
 together with the measured small thermopower cannot be explained by a single band model and is direct evidence of
 two-band conduction with carriers of opposite sign in both bands.

Over the years Greene and coworkers   have performed extensive measurements \cite{greene1,greene2,greene3,greene4,greene5}  on Hall coefficient, thermopower,
magnetoresistance and Nerst effect in electron-doped cuprates, and carefully analyzed their data. 
They found compelling evidence from these measurements for two-band conduction, an electron-band and a hole-band,
and  dominant role of hole carriers in the regime where the materials are superconducting, and in particular that the regime where hole transport begins to dominate coincides with the onset of superconductivity. We believe that these together with other transport experiments and analysis \cite{wang,suzuki,gollnik,greven}
have established experimentally  that 
``{\it in electron doped cuprates holes are responsible for the
superconductivity}''  \cite{greene5}.

\section{quantum oscillations}
In 2007, Shubnikov-de Haas oscillations  in underdoped hole-doped cuprates provided clear evidence for the existence of closed small Fermi surface pockets 
in those materials \cite{dhvah1,dhvah2}. 
This was very surprising in view of the consensus that existed that Fermi liquid concepts did not apply to underdoped cuprates, and that the Fermi surface
consisted of Fermi arcs rather than closed surfaces. The frequency of these oscillations was $530T$
in $YBa_2Cu_33O_{6.5}$ \cite{dhvah1}, corresponding to
$1.9\%$ of the two-dimensional Brillouin zone area, and  $660T$ in $YBa_2Cu_4O_8$ \cite{dhvah2,dhvah3}  with $T_c=80K$, corresponding to
$2.4\%$ of the  Brillouin zone. For the first case $T_c=57.5K$ and $p=0.10$ holes per planar Cu atom, for the second case
$T_c=80K$ and $p=0.125$ \cite{dhvah2} or $p=0.14$ \cite{dhvah3} holes per Cu atom. 
Quantum oscillations have also been observed in underdoped $HgBa_2CuO_{4+\delta}$ \cite{dhvah4,dhvah5}, with frequency 
$840T$ corresponding to about $3\%$ of the Brillouin zone, or 0.061 carriers per pocket.
Several other such observations in hole-doped cuprates have been reported \cite{dhvah6,dhvah7,dhvah8}.
Whether these oscillations are due to hole pockets or electron pockets or both has been controversial 
\cite{hdopedepock,dhvah9}
and is unsettled. 
Quantum oscillations have also been observed in the overdoped regime
of $Tl_2Ba_2CuO_{6+\delta}$  \cite{dhvah10},
with frequency 18,100T, corresponding to cross-sectional area $65\%$ of the Brillouin zone.

Within our model,   in hole-doped cuprates holes are doped into the
$O p\pi$ band that is initially full, so we would expect small hole orbits around the 
$(\pi,\pi)$ point in the Brillouin zone, consistent with the Shubnikov-de Haas  oscillations observed.
It is not clear how these observations can be consistent with doping a half-filled Hubbard band with
holes, since according to what we reviewed in the previous section the Hall coefficient at low
temperatures in that model reflects the bare band structure.

In electron-doped cuprates,  Shubnikov-de Haas oscillations  first detected in 2009 
in $Nd_{2-x} Ce_x CuO_4$ \cite{helm2009} showed that small Fermi surface pockets exist
both in the optimally doped and overdoped samples \cite{helm2009,helm2011,helm2015}. 
The measured frequency is approximately 300T, corresponding to $1.1\%$ of the Brillouin zone. As mentioned earlier,
it is hypothesized that they originate in reconstruction of the Fermi surface, however no long range
antiferromagnetic order exists in this doping range \cite{moto}. 
In addition, from the reconstructed Fermi surface scenario one would expect also quantum oscillations
from electron pockets with 
frequency half of that produced by the hole pockets, however a single  low frequency is observed. 
In $Nd_{2-x} Ce_x CuO_4$ at high doping, both low frequency and high frequency $(\sim 11kT)$ oscillations are detected coexisting
in the range $x>0.15$ to $x=0.17$ \cite{helm2011,helm2015}, with the high frequency oscillations,
corresponding to a Fermi surface area of $\sim 41 \% $ of the Brillouin zone  dominating
at high doping, and the low frequency ones at low doping. The authors suggest that the reconstructed Fermi surface
exists until the point where superconductivity disappears at $x\sim 0.175$ and that the high frequency oscillations
originate in magnetic breakdown orbits,
and they attribute the absence of evidence for electron pockets to damping. They acknowledge that 
``The mechanism responsible for the
broken translational symmetry is still to be clarified'', and suggest `` `hidden'
d-density-wave ordering'' as a possibility. Similar low frequency quantum oscillations are found in
$Pr_{2-x}Ce_xCuO_{4\pm \delta}$ and $La_{2-x}Ce_xCuO_{4\pm \delta}$ \cite{higgins2018}.

Instead, within our model these observations are simply explained by the existence of
a small number of hole carriers in the $O p\pi$ band giving rise to the low frequency oscillations,
and  electrons doped into  the $Cu-O p\sigma$  band giving rise to a large Fermi surface corresponding
to a more than half-filled band and corresponding high frequency Shubnikov-de Haas oscillations.

\section{photoemission}

Recent photoemission investigations on electron-doped cuprates were reviewed by Horio and Fujimori \cite{fuji}.
In particular they point out that the electron concentration inferred from the area of the electron Fermi surface measured by
ARPES is significantly larger than the nominal $Ce$ concentration. This was already found long ago by Alp and coworkers \cite{alp}
via x-ray absorption spectroscopy.
It  is consistent with our picture that
electron doping induces hole carriers in the plane,  hence by charge conservation this generates  extra electron
carriers in addition to the ones doped. 

Horio and Fujimori (HF) emphasize the fact that annealing in a reducing atmosphere plays an essential role in giving rise to
superconductivity, and point out that superconductivity can arise even in the absence of Ce doping.
Contrary to earlier findings, they suggest that reduction removes oxygen from the rare-earth layers rather than from 
impurity oxygens at the apical sites. This is in agreement with our prediction \cite{reduction}.

The photoemission results reviewed by HF show a  Fermi surface developing around $k\sim (\pi,0)$. 
Upon further  doping it becomes a large hole Fermi surface centered around $(\pi,\pi)$.
Similar results are found in photoemission measurements by Song et al \cite{song}. We  infer that all these measurements
are only detecting the quasiparticles residing in the $Cu-O p\sigma$ band, that near half-filling are strongly affected by the Hubbard $U$
 and for sufficient electron doping evolve into a simple Fermi surface centered around $(\pi,\pi)$. These measurements
  don't show the hole carriers that according to the transport measurements must exist in a different band. 
We expect those hole
carriers to be in the $O p\pi$ orbitals forming a small hole pocket at $(\pi,\pi)$, both for electron-doped and 
hole-doped cuprates.
We conclude that because of the strong orbital relaxation effects in this purely oxygen band the quasiparticle weight
is very small \cite{qpw1,qpw2} and not visible in current photoemission experiments.

\section{tunneling asymmetry}

Tunneling measurements have often been at odds with photoemission experiments; both types of experiments have
unknown uncertainties because of the difficulties with surface preparation, etc. as discussed in 
Ref. [{\onlinecite{armitage2010}]. Here we wish to focus on tunneling asymmetry, i.e. the difference in coherence peak
heights, depending on whether the sample is negatively or positively biased with respect to the tip. For hole-doped cuprates, 
a large number of studies have found
that tunneling spectra are asymmetric, with asymmetry of universal sign \cite{tunnhdoped}. For electron-doped cuprates, much less tunneling work has been done. 
We will focus on data
from Refs. [\onlinecite{shan2005,shan2008,shan20082}], since specific details have been clarified through private communication. In particular
in the raw data (Figs. 2 and 3 in Ref. [\onlinecite{shan2005}]) the coherence peak is clearly higher when the sample is negatively
biased with respect to the (normal) tunneling tip. The authors report (private communication) that this is universally true
for all their measurements. Furthermore, this remains true for both Nd$_{1.85}$Ce$_{0.15}$CuO$_{4-y}$ at optimal
doping (subject of Ref. [\onlinecite{shan2005}]) and for Pr$_{1-x}$Ce$_{x}$CuO$_{4-y}$ as a function of doping, 
as shown in Ref. [\onlinecite{shan2008}] (see their Fig.~1) and Ref. \cite{shan20082} (Figs. 1 and 2).

Tunneling spectra of Jubileo and coworkers \cite{jubileo} on $Pr_{1-x}LaCe_xCuO_{4-y}$ also show
a clear asymmetry of the same sign (Figs.1 and  2), and this is also seen in
tunneling on $Pr_{1-x}LaCe_xCuO_{4}$  by Miyakawa et al \cite{miyakawa} (Figs. 1 and 3)
and Diamant et al \cite{diamant}  (Fig. 1).

This tunneling asymmetry is important to confirm since the hole mechanism of superconductivity  \cite{hole1989,hole1990}
predicts an energy-dependent superconducting order parameter that results in a tunneling asymmetry \cite{tunn} of
universal sign, as observed in these experiments. Instead, the RVB model of Anderson and coworkers also predicts
tunneling asymmetry \cite{andersonong} but of opposite sign for hole-doped and electron-doped cuprates
\cite{andersonpriv}. Thus, establishing the sign
of the tunneling asymmetry in electron-doped cuprates can rule out one of these two theories \cite{mareltunn}.

In the theory  of hole superconductivity for a single band, the asymmetry
predicted by this model is clear \cite{tunn}. We will provide further model calculations in the next section, particularly in light of
the fact that the tunneling data given in the different references give no indication of two-band behavior.

\section{model calculations}
We consider a tight binding model for the orbitals shown in Fig. 2. There are 5 orbitals per unit cell $CuO_2$: one for the $Cu$ atom,
and two for each of the oxygens. 
We denote the $d-p\sigma$ hopping amplitude by $t_d$, and the direct hopping amplitudes between oxygen orbitals by $t_1$ for $\pi-\pi$ or $\sigma-\sigma$ hopping
and $t_2$ for $\pi-\sigma$ hopping. Following estimates by McMahan et al \cite{mms} and Stechel and Jennison \cite{sj}
we take $t_1=0.65$, $t_2=0.35$ and $t_d=1.75$, all in $eV$. For site energies
we take $\epsilon_d=-5.2$, $\epsilon_{p\sigma}=-5.5$, $\epsilon_{p\pi}=-4.7$ $eV$. Because of electrostatics, $\epsilon_{p\pi}$ is higher than $\epsilon_{p\sigma}$.
The resulting 5 bands are shown in Fig. 5.
 
            \begin{figure}
 \resizebox{8.5cm}{!}{\includegraphics[width=8cm]{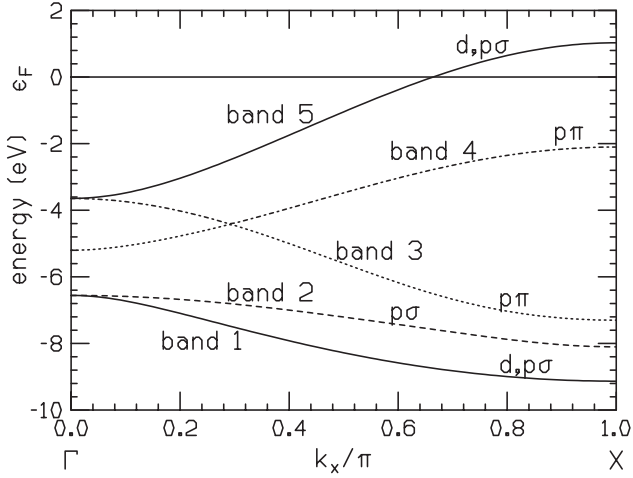}}
 \caption {Band structure in the $Cu-O$ planes in the $\Gamma-X$ direction 
($(0,0)$  to $(\pi,\pi)$) from a tight binding calculation with 5 orbitals per unit cell (see Fig. 2). Parameters used are
$t_{\pi\pi}=t_{\sigma\sigma}\equiv t_1=0.65$, $t_{\pi\sigma}\equiv t_2=0.35$, $t_{d\sigma}\equiv t_d=1.75$,   $\epsilon_d=-5.2$, $\epsilon_{p\sigma}=-5.5$, $\epsilon_{p\pi}=-4.7$   .}.
 \label{figure1}
 \end{figure}
 
 The bands of interest for us are bands 4 and 5. Figure 6 shows the weight of the atomic orbitals
 for the Bloch states of these bands. It can be seen that band 5 has similar content of
 $Cu-d$ orbital and $O p\sigma$ orbital. This is the band that is generally considered to be 
 the important one for both hole-doped and electron-doped cuprates. The Hubbard $U$ opens up a gap  in the undoped 
 compound in this band, rendering the system insulating. From this single band it is
 argued that both hole-like and electron-like carriers can result in the electron-doped cuprates according to the theories discussed in 
 Sect. II.
 
              \begin{figure}
 \resizebox{9.0cm}{!}{\includegraphics[width=8cm]{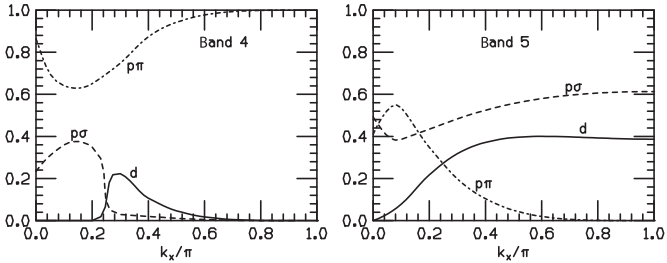}}
 \caption {Weights of the different orbitals in the band states for bands 4 and 5.}
 \label{figure1}
 \end{figure}
 
 Instead, in our picture band 4 is the more important one. As seen in Fig. 6, it is predominantly
 of $O p\pi$ character, particularly as $k$ approaches the $(\pi,\pi)$ point in the 
 Brillouin zone, when the band becomes full. Band structure calculations
 predict that it is about 1.5 eV below the Fermi energy and remains full under doping. Instead, we have argued  \cite{cuo} that strong
 oxygen orbital relaxation makes it easier to create holes in this band than what band structure predicts: qualitatively, when one electron is removed from the doubly occupied $p\pi$ orbital
 the remaining electron orbital shrinks, thus lowering its energy. This local effect is not captured by
 band structure calculations.
 
Band 4 contains the hole carriers that we believe are responsible for superconductivity
 both in hole-doped and electron-doped cuprates. In the hole-doped cuprates, we argue that
 when holes are added to the undoped system they go into this band rather than into band 5. For electron-doped
 cuprates, we argue that adding electrons creates electron carriers in band 5 and 
 through the process depicted in Fig. 3 also creates holes in band 4. We explained how this
 process works in the $FeAs$ compounds in ref. \cite{feas} (Fig. 3) and argue that it is the same here.
 Thus there will be carriers at the Fermi energy both from band 4 and band 5.

We consider a reduced Hamiltonian to describe transport and superconductivity in those 2 bands. Following Suhl et al   \cite{suhl59} we take
\begin{eqnarray}
& & H = \sum_{k\sigma} (\epsilon_{k}^a - \mu)a_{k\sigma}^\dagger a_{k\sigma} +
\sum_{k\sigma} (\epsilon_{k}^d - \epsilon_0 - \mu)d_{k\sigma}^\dagger d_{k\sigma} +
\nonumber \\
& &\sum_{k k^\prime} V_{kk^\prime}^{aa} a_{k\uparrow}^\dagger a_{-k\downarrow}^\dagger
a_{-k^\prime\downarrow} a_{k^\prime\uparrow}
+
\sum_{k k^\prime} V_{kk^\prime}^{dd} d_{k\uparrow}^\dagger d_{-k\downarrow}^\dagger
d_{-k^\prime\downarrow} d_{k^\prime\uparrow}
+
\nonumber \\
& & \sum_{k k^\prime} V_{kk^\prime}^{ad} \bigl( a_{k\uparrow}^\dagger a_{-k\downarrow}^\dagger
d_{-k^\prime\downarrow} d_{k^\prime\uparrow} + d_{k\uparrow}^\dagger d_{-k\downarrow}^\dagger
a_{-k^\prime\downarrow} a_{k^\prime\uparrow}\bigr).
\label{ham}
\end{eqnarray}
As discussed in Ref. \cite{twoband} we retain the simplest interband interaction, and in
what follows adopt a constant interband potential: $V_{kk^\prime}^{ad} = V_{ad}\equiv V_{12}$. We have
used a hole notation, so that the $a^\dagger$ and $d^\dagger$ operators correspond to hole creation
operators in the $O p\pi$ and $Cu-O p\sigma$ band, respectively, and similarly for the annihilation operators.
We adopt a flat density of states for both bands, each with bandwidth $D_i$. The single particle
energies are measured from the center of each band, and the $Cu-O$  band is shifted by an amount
$\epsilon_0$ with respect to the $O$  band.

The intraband potentials are assumed to have identical form; we adopt the form from Ref. \cite{hole1989}:
\begin{equation}
V_{kk^\prime}^{ii}=U_i + K_i \biggl({\epsilon_k^i \over D_i/2} + {\epsilon_{k^\prime}^i \over D_i/2}
\biggr) + W_i {\epsilon_k^i \over D_i/2} {\epsilon_{k^\prime}^i \over D_i/2},
\label{pot}
\end{equation}
where $U_i$ corresponds to the on-site repulsion, $K_i$ the modulated hopping,
and $W_i$ the nearest-neighbor repulsion ($i=1, 2$ correspond to $a, d$ in Eq. (2)). These interactions lead to a BCS ground state that is
superconducting, and an (s-wave) order parameter with the form
\begin{equation}
\Delta_i(\epsilon) = \Delta_i^m\bigl(c_i - {\epsilon \over D_i/2}\bigr),
\label{gap}
\end{equation}
as found previously \cite{twoband,hole1989}. Further details are available in these references.

 Figure 7 shows $T_c$ versus hole concentration in the hole band for two sets of  parameters given in
 the figure caption.  We have taken into account the fact that within our model the
 bandwidth in the hole band increases with hole concentration \cite{m2s89}, i.e. it is not a rigid band model. The behavior 
 shown in Fig. 7 looks similar
to the experimental results in Fig. 4, top right panel. If we were not to take into account the renormalization
of the bandwidth with hole occupation the range of doping where superconductivity occurs would be 
about twice as large as shown in Fig. 7, which would be inconsistent with experiments.

             \begin{figure}[h]
 \resizebox{7.5cm}{!}{\includegraphics[width=6cm]{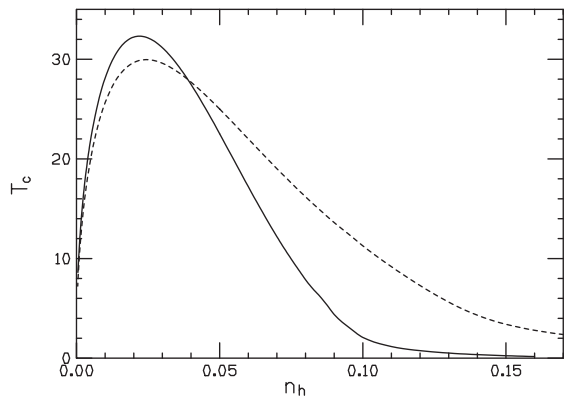}}
 \caption { $T_c$ (K) versus hole  concentration ($n_h$) in our two band model, for parameters
 $D_1=D_2=2$ eV,  $\epsilon_0=1$ eV, $V_{12}=0.8$ eV, $U_1=U_2=10$ eV, $K_2=W_2=0$,  and:
 solid line, $K_1=3.64$ eV, $W_1=0$, dashed line 
$K_1=12.92$ eV, $W_1=16$ eV.} \label{figure1}
 \end{figure}

The fact that the $T_c$ found in experiments (Fig. 4) starts at finite
$Ce$ concentration rather than $0$ would result if  initially the doped electrons and induced holes
remain localized, as discussed in Ref. \cite{reduction} that also suggested an explanation why
oxygen reduction is essential for hole delocalization.

 Fig. 8 shows calculated tunneling spectra for our two-band model  for parameters corresponding
 to the solid line in Fig. 7  just to the right of the maximum with $T_c\sim31K$,
 together compared  with the Shan measurements. For this calculation we used a background density of states of the
 same shape as Shan's, and assumed an intrinsic broadening with
 Dynes' parameter $\Gamma= 0.5 \ {\rm meV}^{-1}$. It can be seen that our results  look very similar
 to Shan's data.
            \begin{figure}[h]
 \resizebox{8.5cm}{!}{\includegraphics[width=5.5cm]{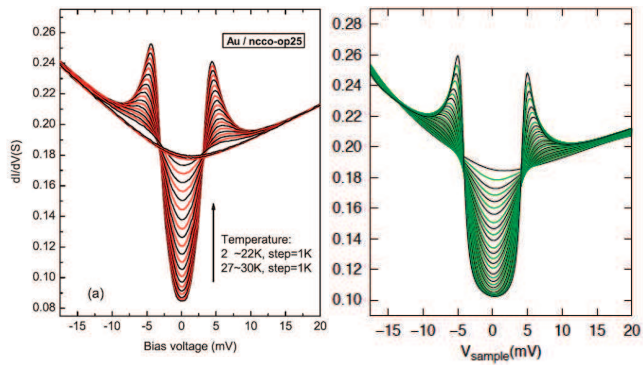}}
 \caption { Comparison of tunneling measurements of Shan et al \cite{shan20082} for a sample with $T_c=25K$ with the predictions
 of the model of hole superconductivity for parameters as in Fig. 7 (solid line)
 with   $n = 0.030$ and $T_c = 31.6$ K.  We have used a
 set of reduced temperatures similar to those used by Shan et al.
The zero temperature gaps are
 $\Delta_1=4.8$ meV, and $\Delta_2=0.05$ meV.  Because of the intrinsic broadening the smaller gap is not visible, even at the lowest
 temperature shown, as is the case in the experimental results.  
 }
 \label{figure1}
 \end{figure}
 
 We conclude that our model is compatible with the experimental tunneling results.

\section{summary and discussion}

In this paper we have argued that a simple two-band model resulting from the orbitals shown in
Fig. 2, where pairing originates in hole carriers in the $O p\pi$ orbitals, can explain in a simple
way a large variety of experimental findings in electron-doped cuprates obtained over many years, as well as in hole-doped cuprates.
The basic principles of the model were proposed before electron-doped cuprates were even discovered.
Our model says that pairing of the same hole carriers drives superconductivity in hole-doped and in electron-doped cuprates.
The pairing mechanism is intimately tied to the hole nature of the carriers and gives rise to high $T_c$ when holes conduct through a
network of negatively charged ions, as we have argued is the case in hole-doped cuprates, electron-doped cuprates,
$MgB_2$,  iron pnictides and chalcogenides and  $H_2S$ \cite{twoband,feas,mgb2,mgb22,h2s}.

In contrast, all other theoretical explanations of experimental observations in electron-doped cuprates assume that  the only electrons
involved are from $Cu$  $d_{x^2-y^2}$ orbitals hybridized with $O$  $p\sigma$ orbitals. To explain how clear experimental signatures of 
two-band physics arise from such a single band necessitates invoking electronic correlations, Fermi surface reconstruction and
hidden exotic orders, which in our view are complicated and contrived explanations not supported by  experimental observations.

In addition to the experiments already discussed, important information about electron-doped cuprates has  been recently  inferred 
from the behaviour of the superfluid density.
The upward curvature of superfluid density versus temperature in electron-doped cuprates has been argued
to be clear evidence for the presence of two types of superfluid carriers \cite{luoxiang}.
Li et al \cite{greven}    analyzed experimental results for upper critical field versus temperature and superfluid density versus temperature in electron-doped cuprates  and concluded that they are consistent with a 2-band model
where $25\%$ of the carriers are hole-like and $75\%$ are electron-like, that the dominant
interaction giving rise to pairing is in the hole-like band, and that  the interband coupling is small
($\lambda_{hh}>>\lambda_{ee}, \lambda_{eh}\sim\lambda_{he}\sim 0$ in their notation). 
Such models have also been used to model two-band superconductivity in $MgB_2$ \cite{nakai}. The superfluid hole 
density inferred by Li et al  \cite{luoxiang} matches the general scaling law between superfluid density and $T_c$ 
proposed by Uemura and coworkers \cite{uemura}.

The  physics uncovered by the analysis of Li et al.  is in agreement with what the model of hole superconductivity
predicts, that was discussed   in Sect. VII and in our earlier work
on hole superconductivity in two-band models \cite{twoband,feas}. In our model, the parameters $K_1$, $K_2$ and $V_{12}$ of Sect. VII are proportional to
$\lambda_{hh}$, $\lambda_{ee}$ and $\lambda_{eh}$ of Li et al \cite{greven}.
In the Li et al. analysis the fact that
holes drive superconductivity is derived from the experimental results. Instead, for us this is a prediction of
the model: when there is two-band conduction with electrons and holes
at the Fermi energy, it is necessarily the holes that pair   and drive the 
entire system superconducting  \cite{twoband}.

Li et al. conclude from their analysis \cite{greven} that it ``{\it points to a single underlying hole-related mechanism of superconductivity in the cuprates regardless of nominal carrier type}''. 
Dagan and Greene concluded from their analysis \cite{greene5}   that {\it ``in electron doped cuprates holes are responsible for the superconductivity''}. Already in 1991 Wang et al had concluded \cite{wang} that
{\it ``The similarity between the behavior
of the hole-scattering rate and  that in earlier ``hole'' superconductors
suggests to us that the holes, in fact, may
be driving the superconducting transition in $Nd_{2-x}Ce_xCuO_{4-\delta}$''}.
These conclusions agree with  what our
model has predicted since 1989. 
To further support  this picture it would be important to confirm experimentally that tunneling asymmetry
{\it of the same sign as for hole-doped cuprates} is the generic behavior in electron-doped cuprates, as initial
experimental results appear to show \cite  {shan2005,shan2008,shan20082,jubileo,miyakawa,diamant}. This would rule out
theories based on electron-hole symmetric models \cite{andersonong}.

There has been conflicting experimental evidence on the question of the symmetry of the order parameter
in electron-doped cuprates  \cite{fournier,armitage2010}. In the optimally doped and overdoped regime there is substantial
evidence of a nodeless order parameter, i.e. of s-wave symmetry \cite{s1,s2,s3}.
Within our model the symmetry of the gap is independent of doping and is the same for electron- and hole-doped
cuprates, namely s-wave. We suggest that the experimental evidence for non-s-wave superconductivity
  in the underdoped regimes of electron-doped and in hole-doped cuprates is due to 
  extrinsic factors, for example the existence of gapless excitations due to strong correlations.

 In summary, our model provides a simple, natural, elegant,  unified and compelling explanation for a wide variety of
 experimental results gathered through intensive experimental research and analysis thereof over three decades
\cite{nucker,wang,fortune,crusellas,suzuki,greene1,greene2,gollnik,greene3,greene4,greene5,greven,helm2009,helm2011,helm2015,fuji,song,armitage2010,shan2005,shan2008,shan20082,jubileo,miyakawa,diamant,fournier}. We suggest that the fact that
 we formulated the model  even before electron-doped cuprates were 
 discovered and predicted the conclusions reached through thirty years of  experimental work argues for the validity
 of the model to describe physical reality.   
 \acknowledgements
 
 FM was supported by the Natural Sciences and Engineering Research Council of Canada (NSERC). We are grateful to R. L. Greene, M. Greven, N. Barisic, H. H. Wen, J. Fink,    N. P. Ong and M. B. Maple for   stimulating discussions on their and other's work on this topic.

\end{document}